# Rotationally Commensurate Growth of MoS$_2$ on Epitaxial Graphene

Xiaolong Liu[1†], Itamar Balla[2†], Hadallia Bergeron[2], Gavin P. Campbell[2], Michael J. Bedzyk[1,2,3*],

and Mark C. Hersam[1,2,4,5*]

[1]*Graduate Program in Applied Physics,* [2]*Department of Materials Science and Engineering,*

[3]*Department of Physics,* [4]*Department of Chemistry,* [5]*Department of Medicine*

*Northwestern University, Evanston, IL 60208, USA*

† These authors (X.L. and I.B.) contributed equally.

*Correspondence should be addressed to:

m-hersam@northwestern.edu; bedzyk@northwestern.edu

**ABSTRACT**

Atomically thin MoS$_2$/graphene heterostructures are promising candidates for nanoelectronic and optoelectronic technologies. Among different graphene substrates, epitaxial graphene (EG) on SiC provides several potential advantages for such heterostructures including high electronic quality, tunable substrate coupling, wafer-scale processability, and crystalline ordering that can template commensurate growth. Exploiting these attributes, we demonstrate here the thickness-controlled van der Waals epitaxial growth of MoS$_2$ on EG *via* chemical vapor deposition, giving rise to transfer-free synthesis of a two-dimensional heterostructure with registry between its constituent materials. The rotational commensurability observed between the MoS$_2$ and EG is driven by the energetically favorable alignment of their respective lattices and results in nearly strain-free MoS$_2$, as evidenced by synchrotron X-ray scattering and atomic-resolution scanning tunneling microscopy (STM). The electronic nature of the MoS$_2$/EG heterostructure is elucidated with STM and scanning tunneling spectroscopy, which reveals bias-dependent apparent thickness, band bending, and a reduced bandgap of ~0.4 eV at the monolayer $MoS_2$ edges.

**TOC Figure:**

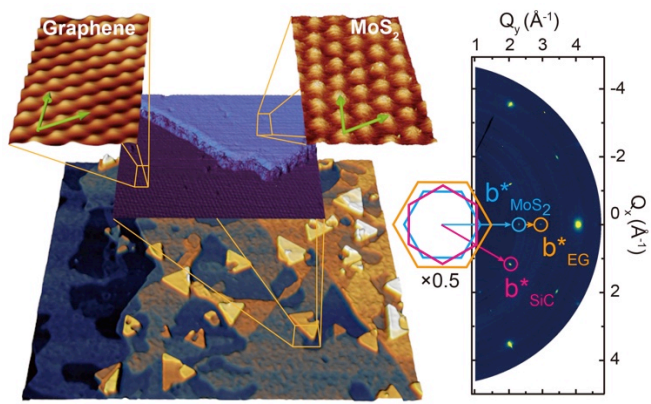

Vertical heterostructures composed of stacked two-dimensional (2D) materials allow the exploration of fundamental interfacial interactions and novel electronic functionality. From among the 2D material library consisting of semimetals (*e.g.*, graphene[1,2]), insulators (*e.g.*, boron nitride[3] and $Bi_2Se_3$[4]), and semiconductors (*e.g.*, black phosphorus[5-7] and transition metal dichalcogenides[8,9]), the $MoS_2$/graphene heterostructure[10,11] shows great potential for next generation electronic and optoelectronic applications due to complementary carrier mobilities and optical responsivities.[12] Additionally, the $MoS_2$/graphene heterostructure has been shown to be an excellent catalyst for hydrogen evolution reactions[13] and a promising anode material for Li-ion batteries.[14]

The properties of $MoS_2$/graphene heterostructures depend strongly on the underlying substrate and the graphene synthesis technique. Epitaxial graphene (EG) grown on SiC by the preferential thermal desorption[15,16] of silicon from SiC not only offers uniform large-area synthesis of graphene, but also provides technological advantages over alternative methods such as chemical vapor deposition (CVD) and mechanical exfoliation. For example, fine control of the growth temperature enables homogeneous monolayer or bilayer EG at the wafer scale,[17-19] while post-annealing in hydrogen allows decoupling of EG from the underlying SiC substrate.[2] The high quality and cleanliness of EG is evidenced by its high carrier mobility[20] of 45,000 cm$^2$V$^{-1}$s$^{-1}$ and observation of the quantum hall effect.[21] Furthermore, the interaction between graphene and the SiC substrate can be tailored *via* intercalation of different atomic species.[22]

Given the underlying SiC substrate as well as the different electronic characteristics of EG (*e.g.*, substrate-induced n-type doping[23,24]), CVD synthesis of MoS$_2$ on EG is expected to differ from that on graphite or CVD graphene, giving rise to novel structural and electronic interactions between the two materials. While MoSe$_2$ has been grown on EG by molecular beam epitaxy,[25,26] recent reports of MoS$_2$ grown by CVD on graphite[9,27,28] and EG[29,30] resulted in non-epitaxial growth, where the orientation of the MoS$_2$ crystals was not controlled. Furthermore, the atomic-scale electronic and structural properties of MoS$_2$/EG heterostructures have not been thoroughly established. Towards these ends, we report here the rotationally commensurate growth of atomically thin MoS$_2$ crystals on EG by van der Waals epitaxy. We investigate the electronic and structural properties of the resulting MoS$_2$/EG heterostructures by a comprehensive suite of methods including atomic force microscopy (AFM), ultra-high vacuum (UHV) scanning tunneling microscopy and spectroscopy (STM/STS), synchrotron X-ray scattering, *in situ* X-ray photoelectron spectroscopy (XPS), and Raman spectroscopy. In addition to the inherent advantages of van der Waals epitaxy (*e.g.*, reduced defect density,[31] sharper interfaces,[32] and consistent material quality), the registry between the MoS$_2$ and EG allows for the study of crystal orientation-dependent properties (*e.g.*, anisotropic thermal conductance of MoS$_2$[33]) and specific types of grain boundaries (*e.g.*, mirror twin grain boundaries[34]) for fundamental studies[34] and electronic applications.[35] Specifically, the epitaxial growth of MoS$_2$ hinders the formation of tilt grain boundaries that degrade in-plane electrical conductivity, and favors the formation of mirror twin grain boundaries that preserve in-plane electrical conductivity.[34] It has also been demonstrated that the type of grain boundary affects the optical properties of MoS$_2$.[34] Therefore, the ability to synthesize rotationally commensurate MoS$_2$ domains on EG facilitates ongoing efforts to realize reproducibly high performance MoS$_2$/graphene electronic and optoelectronic devices.

**RESULTS AND DISCUSSION**

CVD growth of MoS$_2$ on EG is performed at a variety of conditions in order to tune and optimize the MoS$_2$ film thickness. Figure 1a is an STM image of EG prior to MoS$_2$ growth that contains a rotational grain boundary on EG with atomic structure consistent with literature precedent.[36,37] In addition to the rotational grain boundary defect, the graphene lattice and underlying (6√3×6√3)R30° SiC reconstruction are clearly resolved. Figures 1b and 1c are AFM topography and phase images, respectively, of MoS$_2$ grown on EG at 43 Torr (detailed growth conditions are given in Methods; additional AFM images are provided in Figure S1). The measured thickness of a MoS$_2$ triangular domain is 7.0 Å as indicated by the green line profile in Figure 1b, corresponding to monolayer MoS$_2$.[38] Monolayer and bilayer graphene regions are observed in the topography image (Figure 1b), and are more clearly distinguished in the phase image (Figure 1c), where monolayer regions are brighter. The measured step height is 3.5 Å as indicated by the red line profile in Figure 1b, which is also in agreement with literature.[39] Figure 1d extracts the edge orientations of the MoS$_2$ triangles shown in Figure 1b by plotting a line along one edge of each triangle. A large majority of the triangles have parallel edges as indicated by the blue lines, while a few triangles, indicated in red, have edges rotated by 30°. The rare occurrence of a third rotational orientation of a MoS$_2$ triangular domain is shown in pink. From this analysis, it is apparent that there are two predominant types of azimuthal registration of MoS$_2$ on EG. This crystal orientation alignment is attributed to van der Waals epitaxy,[32] which accommodates the large lattice mismatch (~28%) between graphene and MoS$_2$, as will be further discussed below.

Examination of the growth nucleation sites reveals that MoS$_2$ growth preferentially (but not exclusively) initiates at the step edges of the underlying SiC and graphene. This observation can be explained by the higher reactivity of graphene as a function of local curvature[40,41] at SiC steps and exposed graphene edges.[42] Indeed, STS spatial maps reveal different electronic properties of the graphene wrinkles and edges compared to the flat regions as shown in Figure S2. Higher growth pressures of 50 Torr and 100 Torr result in bilayer and pyramid-like multilayer growth with larger domain sizes and higher coverage as shown in Figure 1e and 1f, respectively, where the interface between EG and MoS$_2$ is highlighted in red. Higher synthesis pressure increases the local concentration of growth precursors and facilitates growth on the MoS$_2$ basal planes, resulting in multilayer crystals.[43,44] Hence, by tuning the growth pressure, the thickness of the MoS$_2$ domains can be controlled.

The as-grown MoS$_2$/EG heterostructure is characterized by Raman spectroscopy as shown in Figure 2a. Raman signatures of MoS$_2$ (A$_{1g}$ at 407.2 cm$^{-1}$, E$^1_{2g}$ at 386.4 cm$^{-1}$) and graphene (D at 1391.2 cm$^{-1}$, G at 1575.4 cm$^{-1}$, 2D at 2740.8 cm$^{-1}$) are clearly resolved. Several factors affect the positions of the A$_{1g}$ and E$^1_{2g}$ modes including layer number,[45] strain,[46] and doping.[47] Using a method previously demonstrated on graphene,[48] the strain and doping effects of MoS$_2$ on EG can be deconvolved. For this analysis, a sample of CVD-grown MoS$_2$ on SiO$_2$ was prepared (see the optical microscopy and AFM images in Figure S3). To circumvent MoS$_2$ thickness effects on the Raman peak positions, the two samples were prepared under growth conditions that yield primarily MoS$_2$ monolayers. Spatial Raman mapping of the MoS$_2$/EG and MoS$_2$/SiO$_2$ samples was then performed, and the E$^1_{2g}$ *versus* A$_{1g}$ mode positions are plotted in Figure 2b. It has been shown previously that A$_{1g}$ and E$^1_{2g}$ modes shift up with compressive strain[46] and down with tensile strain.[49,50] Since the reported ratio of the shifting rates of E$^1_{2g}$ to A$_{1g}$ under biaxial strain is ~1.5[46] and hole doping shifts up the A$_{1g}$ mode while the E$^1_{2g}$ mode remains unchanged,[47] pure strain and doping effects are represented by two lines with slopes of 1.5 and 0 in Figure 2b, respectively. Thus, in comparison to MoS$_2$ on SiO$_2$, MoS$_2$ grown on EG has higher hole doping and lower tensile strain. The higher hole doping level may be due to differences in charge transfer for the two systems.[51,52] In addition, the lower tensile strain can be explained by the fact that while MoS$_2$/SiO$_2$ is under tensile strain,[53] MoS$_2$/EG is strain-free as will be demonstrated later by X-ray scattering. Given that the difference in the E$^1_{2g}$ mode positions between the two systems is 2.4 cm$^{-1}$, the corresponding tensile strain of MoS$_2$/SiO$_2$ is ~0.2 %.[46]

The electronic properties of the MoS$_2$/EG heterostructure are further explored using a custom-designed UHV STM with a base pressure of ~6×10$^{-11}$ Torr at room temperature.[54] Prior to scanning, the sample is degassed at 205 °C for 6 hours in UHV and then characterized by *in situ* XPS. The molybdenum core level XPS spectrum is shown in Figure 2c with detailed subpeak positions provided in the Methods section. The doublets of the Mo 3d-orbital corresponding to MoS$_2$ are considerably higher than those of MoO$_x$, indicating the high quality of the MoS$_2$ crystals. Figure 3a is an STM image showing a typical monolayer MoS$_2$ crystal on EG. The (6√3×6√3)R30° reconstruction of EG is seen in the bottom half of the image. Nine STS spectra taken at different positions on the MoS$_2$ crystal far from its edges are shown in Figure 3b. The bandgap is uniformly ~2 eV across a single domain, which falls within the range of reported STS bandgaps of monolayer MoS$_2$ on graphite (1.9 eV[28], 2.15 eV[9] and 2.4 eV[27]). Thermal broadening and energy uncertainty resulting from the small tunneling junction may explain the small deviations in the STS-measured bandgap.[55] Atomic resolution images of a MoS$_2$ triangular domain and EG in Figure 3a are shown in Figures 3c and 3d, respectively. From the line profiles, the periods of atomic corrugation for MoS$_2$ and EG are measured to be 3.15 Å and 2.46 Å, corresponding to their respective lattice constants. Both the MoS$_2$ and EG atomic-scale STM images show 6-fold symmetry and rotational commensurability as a result of van der Waals epitaxy. Figure 4a contains STS mappings of the corner of a MoS$_2$ triangle overlaid on three-dimensionally rendered STM height images taken at V$_{sample}$ = -0.3 V and V$_{sample}$ = 0.3 V. Since the differential tunneling conductance measured by STS is proportional to the sample density of states (DOS), the semiconducting nature of MoS$_2$ and semimetallic nature of EG give rise to the inversion of the relative DOS at these two biases.

STS spectra for monolayer and bilayer MoS$_2$ are shown in Figure 4b. Compared to monolayer, bilayer MoS$_2$ has a reduced bandgap of ~1.7 eV resulting from an increased valence band maximum (VBM) and a subtle decrease of the conduction band minimum (CBM). Although the physical thickness of monolayer MoS$_2$ is ~7 Å, the apparent height measured by STM depends on applied bias, as indicated by the blue curve in Figure 4c. This convolution of topography and electronic structure in constant current STM imaging is well-known and often pronounced for ultrathin films.[56] Qualitatively, at high positive bias, the MoS$_2$ conduction band shows higher DOS than EG as shown in Figure 4a, which implies that the tip retracts more on MoS$_2$ than EG to maintain a constant current. On the other hand, at low negative bias, the opposite occurs as EG shows higher DOS than MoS$_2$ (Figure 4a). This change in relative DOS and the proximity of the MoS$_2$ CBM to zero makes the MoS$_2$ apparent thickness larger at high positive bias and smaller at negative bias than its physical thickness. Also shown in Figure 4c is the measured apparent height of the step edge from bilayer to monolayer MoS$_2$ at different sample biases. Since bilayer and monolayer MoS$_2$ have similar DOS in the measured range, the apparent thickness does not vary significantly from its physical thickness.

The effects of edges[57,58] on the electronic properties of MoS$_2$ are also investigated by spatially resolved STS spectra. A series of vertically offset STS spectra taken across an EG to MoS$_2$ step edge are shown in Figure 5a, with their respective positions marked by colored dots in the inset. The distance from each measured point to the step edge is further indicated in the figure. These spectra reveal that the bandgap increases gradually to ~2 eV after moving ~5 nm into the MoS$_2$ flake. The dashed black lines denote the VBM and CBM of each spectrum. Both band edges shift to higher energy when approaching the MoS$_2$ edge from the interior, similar to the band bending observed for MoS$_2$ on graphite.[9] However, the gradual decrease of the bandgap (*i.e.*, faster increase of the VBM than the CBM) when approaching the MoS$_2$ edge differs from the nearly constant bandgap observed in the band bending region of MoS$_2$ on graphite.[9] Similar STS spectra taken across the transition from bilayer to monolayer MoS$_2$ are shown in Figure 5b. In this case, the bandgap changes from ~1.7 eV on bilayer MoS$_2$ to ~2.0 eV on monolayer MoS$_2$. However, at the edge position, the bandgap drops significantly to ~0.4 eV. This large decrease in the bandgap is consistent with edge states[58] that have been predicted by density function theory (DFT) to reduce the bandgap of MoS$_2$ to 0.43 eV at the monolayer to bilayer step edge[59] and below 0.6 eV at isolated MoS$_2$ monolayer edges.[57]

To verify that these local AFM and STM observations persist over the entire sample, synchrotron grazing-incidence wide-angle X-ray scattering (GIWAXS) is employed to investigate the van der Waals epitaxy of MoS$_2$ on EG. A schematic of GIWAXS is shown in Figure 6a and experimental details are outlined in the Methods section. In GIWAXS, the incident X-ray beam is at incident angle $\alpha$ = 0.14°, slightly below the critical angle of total external reflection of the substrate. The out-of-plane angle $\beta$ and in-plane angle $2\theta$ define the position at which the scattered X-ray wave vector $k_f$ intersects the plane of the 2D detector. The $Q_{xy}$ 2D reciprocal space map of the MoS$_2$/EG/SiC structure projected from $Q_z$ = 0.08 Å$^{-1}$ to 0.12 Å$^{-1}$ is shown in Figure 6b with first order spots of MoS$_2$, EG, and SiC identified. The blue, orange, and magenta arrows indicate the reciprocal lattice vectors **b**$^*$ for MoS$_2$, EG, and SiC, respectively. As seen by the reciprocal space points, the reciprocal lattices for MoS$_2$ and EG are aligned, whereas SiC has a well-known 30° rotation with respect to EG.[60] This observation indicates that the majority of the MoS$_2$ crystals are epitaxially grown along the EG lattice direction with matching six-fold symmetry, confirming the symmetry inferred from AFM (Figure 1 and Figure S1) and STM (Figure 3). The real-space structure reconstructed for such aligned MoS$_2$ growth is shown in Figure 6c, where the two MoS$_2$ triangles represent the preferred orientations in the epitaxial MoS$_2$/EG heterostructure shown in Figures 3c and 3d. Projected first order MoS$_2$ and SiC peaks onto $\phi$ are included in Figure 6d to examine the angular distribution. The relative angle between the MoS$_2$ and SiC lattices is indeed 30° ± 0.3°, which is a much narrower azimuthal distribution than that of CVD MoS$_2$ grown on sapphire.[61] The full-width-at-half-maxima of the first order $MoS_2$, EG, and SiC peaks as a function of $\phi$ are 0.7°, 0.5°, and 0.05° (limited by X-ray optics), respectively, which confirms that the $MoS_2$ is in excellent registry with EG.

Figure 6e shows the scattered intensity from the 2D reciprocal space map of Fig. 6b collected along $Q_y$ at $Q_x = 0$. The real-space lattice constants calculated for $MoS_2$, EG, and SiC are 3.16 ± 0.01 Å, 2.46 ± 0.01 Å and 3.07 ± 0.01 Å, respectively. To gain more insight into the structure of $MoS_2$ on EG, in-plane X-ray diffraction of $MoS_2$/EG and bulk $MoS_2$ was performed (Figure S4). The resulting lattice constants of $MoS_2$/EG and bulk $MoS_2$ are 3.160 ± 0.005 Å and 3.159 ± 0.006 Å, which indicate a relaxed in-plane structure of $MoS_2$ on EG similar to its bulk counterpart. The lack of in-plane strain in the synthesized $MoS_2$ 2D crystals can be explained by total strain relaxation in the van der Waals gap, analogous to a buffer layer alleviating the strain in conventional epitaxy.[62]

As indicated by the red $MoS_2$ triangles in Figure 1d, there is a secondary, less preferred registration between $MoS_2$ and EG, where the $MoS_2$ lattice is rotated by 30°. To examine the relative amount of 30° rotated growth, GIWAXS data of the $MoS_2$(010) peak taken along the EG[010] direction (aligned growth, blue) and the SiC[010] direction (30° rotated growth, red) are shown in Figure 7. From the integrated intensities of the two $MoS_2$ peaks, the amount of $MoS_2$ with lattice 30° rotated from the EG lattice is determined to be 14 ± 4% of the total intensities while the aligned growth is found to be dominant at 86 ± 4%. This large-area GIWAXS result is consistent with the orientation distribution of $MoS_2$ domains shown in the AFM images in Figures 1 and S1. Thus, in Figure 1d, we attribute the majority blue $MoS_2$ triangles to aligned growth, and the minority red triangles to 30° rotated growth. The inset of Figure 7 shows the schematic of these two registrations, where the blue, black, and red arrows indicate the armchair directions of aligned $MoS_2$, EG, and 30° rotated $MoS_2$, respectively. These two growth orientations have also been observed previously for $MoS_2$ grown on sapphire.[61,63] Rare deviations in the preferred orientations, such as the pink $MoS_2$ crystal domain shown in Figure 1d, indicate lack of registry of the $MoS_2$ domain with the EG and can be attributed to local disturbances in the substrate surface. These occurrences were not frequent enough to produce detectable scattered X-ray intensity.

**CONCLUSIONS**

In summary, a 2D heterostructure of rotationally commensurate $MoS_2$ on EG has been grown *via* van der Waals epitaxy with controlled thickness. The structural and electronic quality of these samples has been probed down to the atomic scale using a combination of UHV STM/STS and synchrotron X-ray scattering. $MoS_2$ is found to preferentially grow with lattice aligned with EG. The relative amount of a less preferred 30° rotated growth is also determined. Furthermore, Raman spectroscopy and in-plane X-ray diffraction show that the $MoS_2$ is nearly strain-free, thus providing an ideal system for probing the fundamental properties of two-dimensional $MoS_2$. For example, spatially resolved STS measurements across $MoS_2$ step edges reveal bandgap narrowing that is in close agreement with DFT calculations. Overall, this study suggests that EG may be a promising substrate for van der Waals epitaxial growth of other emerging two-dimensional nanomaterials in addition to providing a well-defined platform for the future study and application of $MoS_2$/graphene heterostructures.

**METHODS**

**Synthesis of Epitaxial Graphene.** Following a similar procedure of producing EG on 6H-SiC (0001) as outlined previously,[15,16] an n-type 4H-SiC (0001) wafer (Cree Inc.) degreased in an acetone and isopropyl alcohol (IPA) sonication bath was degassed at 550 °C for 12 hr under ultra-high vacuum (UHV) conditions (~$5 \times 10^{-10}$ Torr). Few-layer EG was grown on the Si face by thermal desorption of Si atoms at 1270 °C for 20 min. The first carbon rich buffer layer has a 6√3×6√3 unit cell, which is known as the (6√3×6√3)R30° reconstruction.

**Synthesis of $MoS_2$.** Molybdenum trioxide ($MoO_3$, 99.98% trace metal Sigma Aldrich) was placed in the middle of the hot zone of a Lindberg/Blue 1'' quartz tube furnace in an alumina boat 2.5 cm upstream of a 5 mm × 9 mm EG substrate (graphitized Si face is facing up). The sulfur powder (Sigma-Aldrich) in an alumina boat was placed 30 cm upstream of the $MoO_3$ boat under a proportional-integral-derivative (PID) temperature controlled heating belt. The tube was initially pumped to a base pressure of ~50 mTorr and purged with Ar gas to 400 Torr. During the anneal prior to the reaction and during the reaction itself, the pressure was kept constant at 43 Torr (to yield mostly monolayer $MoS_2$ crystals) using a needle valve controller and Ar carrier gas flowing at 25 sccm. The $MoO_3$ and EG substrate were annealed for 20 min at 150 °C (with a 5 min ramp to 150 °C from room temperature) to eliminate residual water and physisorbed contaminants in the tube and on the substrate. Subsequently, the furnace was heated to a maximum temperature of 800 °C at a rate of 12 °C/min. Once the target temperature was reached, the furnace was kept at 800 °C for 20 min and cooled down naturally to room temperature. Concurrently, the sulfur was annealed under the same inert conditions for 49 min at 50 °C (with 5 min ramp to 50 °C from room temperature) and brought to a maximum temperature of 140 °C at a rate of 4 °C/min. The sulfur ramp to 140 °C began when the furnace was approximately at 500 °C. The sulfur was then kept at the maximum temperature for 23 min.

For the synthesis of $MoS_2$ on $SiO_2$, a 4 cm × 1 cm 300 nm thick $SiO_2$/Si wafer (SQI Inc.) was sonicated in acetone and IPA bath for 10 min, rinsed with deionized water, and treated with oxygen plasma for 1 min. The wafer was then placed in close proximity to the $MoO_3$ boat and the same growth procedure was applied with a change to the following parameters: a reaction chamber pressure of 150 Torr was maintained instead of 43 Torr and the sulfur was brought to a maximum temperature of 150 °C instead of 140 °C at a rate of 4.5 °C/min.

**Scanning Tunneling Microscopy and Spectroscopy.** A home-built UHV STM system[54] integrated with a preparation chamber, an XPS chamber, and a load lock was utilized for STM, STS, and XPS measurements. The microscope adopts a Lyding design[64] with bias voltage applied to the sample with respect to the grounded tip. The base pressure in the STM chamber is $\sim 6 \times 10^{-11}$ Torr. Both electrochemically etched W and PtIr tips (Keysight) were used. STS spectra and mappings were taken with a lock-in amplifier (SRS Model SR850) with a modulation frequency of ~8.5 kHz and RMS amplitude of 30 mV. The experimental variability of bandgap measurements in this work is estimated to be 0.08 eV. This estimate is based on the standard deviation of bandgap measurements (Figure S5) from 80 STS spectra on multiple monolayer $MoS_2$ domains. Specifically, these measurements imply that the bandgap is 2.01 ± 0.08 eV. Nanonis SPM control electronics were used for data collection. The STM system was calibrated using atomic resolution EG for the x-y piezo and Ag(111) single crystal monolayer step height for the z piezo.

**Atomic Force Microscopy.** AFM experiments were carried out on an Asylum Cypher AFM in tapping and contact modes. Si cantilevers NCHR-W from NanoWorld with resonant frequency of ~320 kHz were used for tapping mode imaging, and PPP-CONTSCAu cantilevers from NanoSensors were used for contact mode imaging (2-5 nN force was applied). The images were taken with a pixel resolution of 512 × 512 or 1024 × 1024 at a scanning rate of ~1 Hz.

**X-Ray Photoelectron Spectroscopy.** XPS spectra were taken at a base pressure of ~4 × $10^{-10}$ Torr with an Omicron DAR 400M X-ray source, XM 500 X-ray monochromator, and EA 125 energy analyzer. The XPS system is integrated with the UHV STM system for *in situ* measurements. The binding energy resolution was 0.1 eV. Ten scans were averaged for each core level spectrum. All subpeaks were fit with a modified Shirley background in Avantage (Thermo Scientific) software after calibrating the spectra to graphene carbon at 284.5 eV. The peaks positions are: S 2s: 227.0 eV, $MoS_2$ $3d_{5/2}$: 229.7 eV, $MoS_2$ $3d_{3/2}$: 232.8 eV, $MoO_x$ $3d_{5/2}$: 233.1 eV, and $MoO_x$ $3d_{3/2}$: 236.3 eV.

**Raman Spectroscopy.** Raman measurements were carried out on a Horiba Scientific XploRA PLUS Raman microscope with excitation laser line of 532 nm (with spot size ~1 $\mu m^2$) in ambient conditions. The Raman signal was collected using a 100X Olympus objective (NA = 0.9) and dispersed by 1800 grooves/mm grating to a Syncerity CCD detector with a corresponding spectral resolution finer than 2 $cm^{-1}$. A laser power of about 1 mW was used with an acquisition time of ~15 s to avoid heating effects. The Raman spectra were collected in the spatial mapping mode with step sizes of ~500 nm. The $MoS_2$ Raman modes of $E^1_{2g}$ and $A_{1g}$ on both the EG/4H-SiC and $SiO_2$/Si substrates were fitted with Lorentzian functions and calibrated against the 4H-SiC band[65] at 776.65 $cm^{-1}$ and Si band at 520.7 $cm^{-1}$, respectively.

**Synchrotron X-Ray Scattering.** The GIWAXS data of $MoS_2$/EG/SiC were collected at sector 12-ID-C at the Argonne National Laboratory Advanced Photon Source. The 23.5 keV monochromatic X-ray beam was defined using slits of 50 μm × 2 mm to have a 5 mm × 2 mm footprint on the sample at incident angle α = 0.14°. The incident flux was 4×$10^{12}$ photons/s. As depicted in Figure 6a, a 100K Pilatus Area Detector was mounted on a rotating ν-arm positioned at approximately 354 mm from the sample to collect the GIWAXS signal. Data within q = 2.2 $Å^{-1}$ to 4.8 $Å^{-1}$ in Figure 6 was collected at ν = 10.262° relative to the direct beam. The sample was kept under a helium environment and placed on a ϕ rotation stage. The 33-ID-C line was used to collect high-resolution grazing-incidence X-ray scattering and diffraction data from the bulk $MoS_2$ single crystal and $MoS_2$/EG samples. Monochromatic 10 keV X-rays were focused to 70 μm × 30 μm using a horizontal Kirkpatrick-Baez mirror with a flux of 2×$10^{11}$ photons/s. Data were collected using a Dectris 100K Pilatus area detector mounted on a Newport 6-circle goniometer. The reciprocal space map in Figure 6b was generated using the Ewald sphere construction method, wherein pixels from $Q_z = 0.08$ Å$^{-1}$ to $0.12$ Å$^{-1}$ were projected onto a 2D reciprocal space map using interpolation plots in Mathematica. Peaks were fit using the Gaussian fitting function in MATLAB.

*Conflict of Interest:* The authors declare no competing financial interests.

*Supporting Information Available:* Additional information including supplemental AFM, STM/STS, optical images, and lattice constants analysis. This material is available free of charge *via* the Internet at http://pubs.acs.org.

*Acknowledgment.* CVD growth was supported by the National Institute of Standards and Technology (NIST CHiMaD 70NANB14H012), STM/STS characterization was supported by the U.S. Department of Energy SISGR program (DOE DE-FG02-09ER16109), Raman and XPS characterization was supported by the Office of Naval Research (ONR N00014-14-1-0669), and synchrotron X-ray scattering measurements were supported by the Materials Research Science and Engineering Center (MRSEC) of Northwestern University (NSF DMR-1121262). Use of the Advanced Photon Source at Argonne National Laboratory was supported by DOE-BES (DE-AC02-06CH11357). The Raman instrumentation was funded by the Argonne-Northwestern Solar Energy Research (ANSER) Energy Frontier Research Center (DOE DE-SC0001059). The authors kindly thank Dr. Kan-Sheng Chen, Dr. Jian Zhu, Dr. Junmo Kang, Dr. Joshua Wood, Dr. Jonathan Emery, Dr. Zhan Zhang, Dr. Hua Zhou, Andrew Mannix, and Zonghui Wei for valuable discussions.

**Figures**

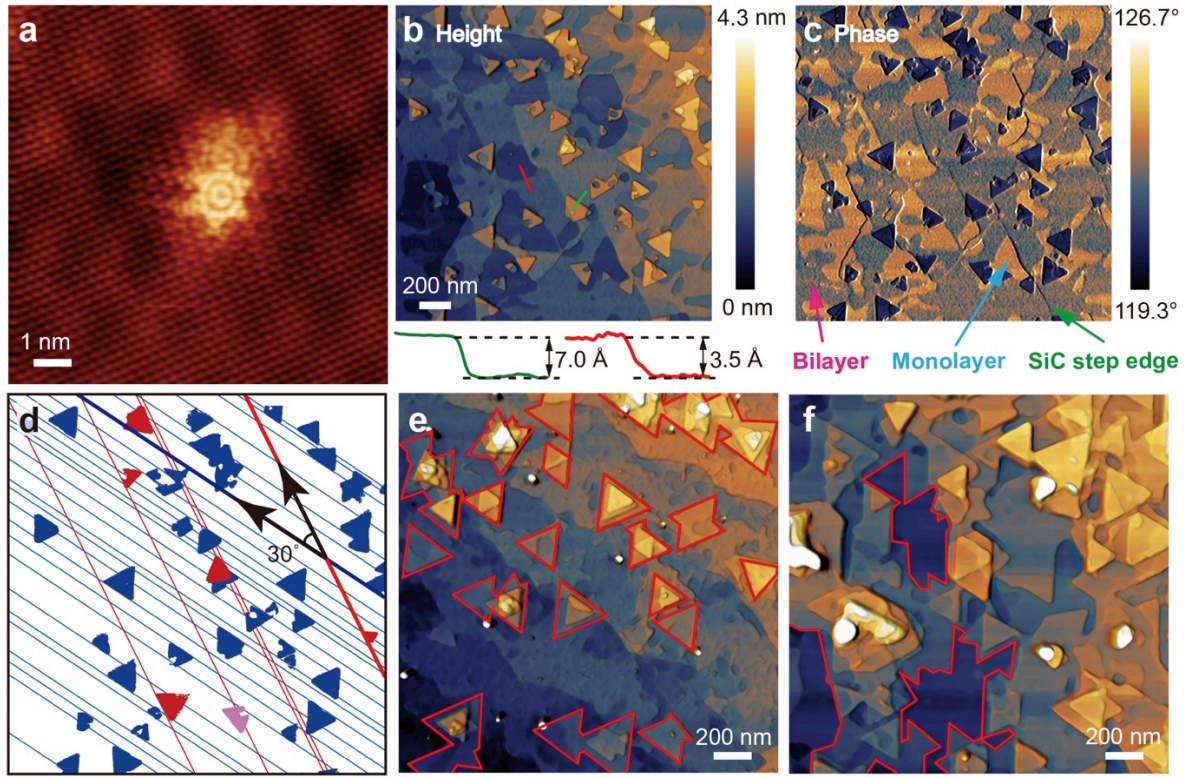

**Figure 1.** CVD-grown MoS$_2$ on EG at different conditions. (a) UHV STM image of EG before growth of MoS$_2$ showing a typical rotational grain boundary on EG (V$_{sample}$ = -0.1 V, I$_{tunneling}$ = 0.4 nA). EG lattice and underlying SiC (6√3×6√3)R30° reconstruction are clearly observed. (b) AFM height and (c) phase images of MoS$_2$/EG grown at 43 Torr. Line profiles show monolayer MoS$_2$ (green) and graphene (red) thicknesses. The contrast between bilayer and monolayer regions of EG is more obvious in the phase image. (d) Extraction of edge orientations of MoS$_2$ crystals in (b) showing two predominant registrations of MoS$_2$ on EG. (e) AFM images of MoS$_2$/EG grown at 50 Torr and (f) 100 Torr showing multilayer growth and larger crystal domain size. The interface between EG and MoS$_2$ is highlighted in red.

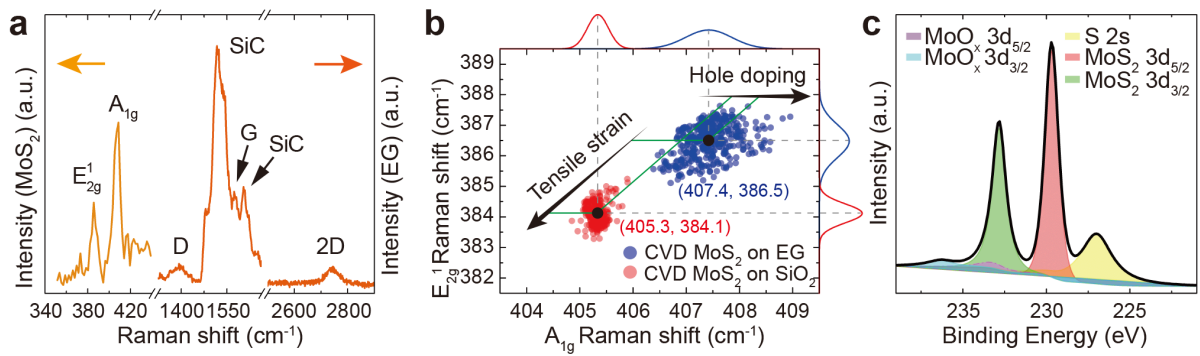

**Figure 2.** Raman and XPS characterization of a MoS$_2$/EG heterostructure with primarily monolayer MoS$_2$. (a) Raman spectrum of MoS$_2$/EG showing the E$^1_{2g}$ and A$_{1g}$ peaks of MoS$_2$, and the D, G, and 2D peaks of EG. (b) Strain and doping analysis of CVD MoS$_2$ on EG in comparison to CVD MoS$_2$ on SiO$_2$ from spatial Raman mapping measurements. (c) *In situ* XPS spectrum of MoS$_2$/EG after annealing at 205 °C for 6 hours in UHV showing characteristic Mo 3d doublets and minimal MoO$_x$ peaks.

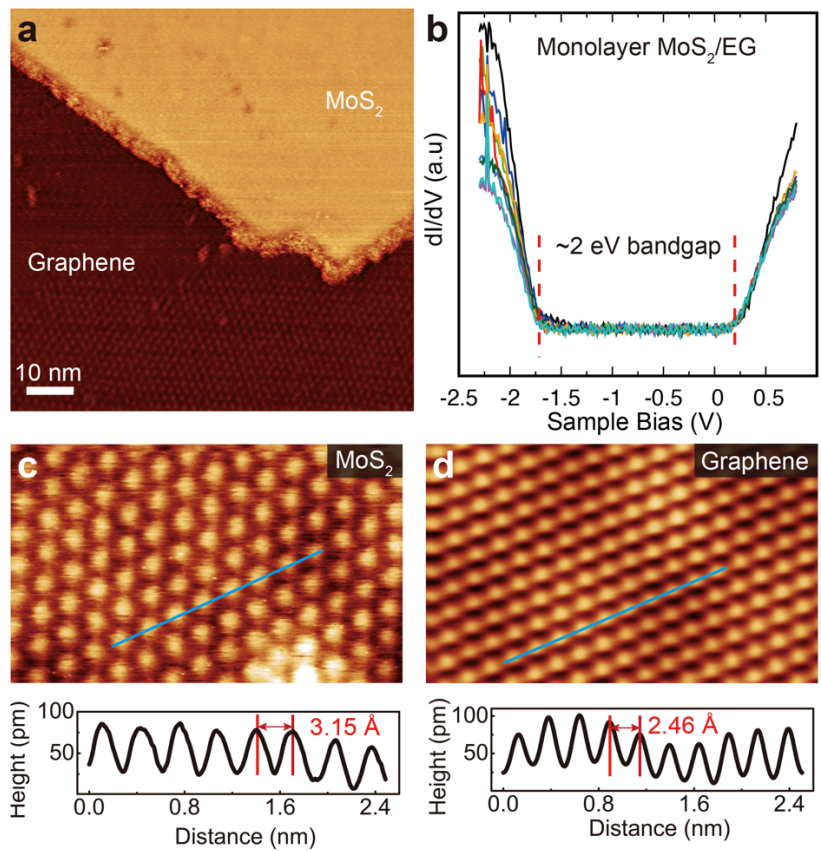

**Figure 3.** UHV STM/STS characterization of MoS$_2$/EG. (a) STM image of the corner of a monolayer MoS$_2$ crystal domain on EG, where the underlying SiC (6√3×6√3)R30° reconstruction is seen at the bottom (V$_{sample}$ = 0.2 V, I$_{tunneling}$ = 50 pA). (b) STS dI/dV spectra taken at 9 different positions on the MoS$_2$ crystal domain far away from the edges showing a uniform bandgap of ~2 eV. (c,d) Atomic-scale images and line profiles of (c) MoS$_2$ (V$_{sample}$ = -0.2 V, I$_{tunneling}$ = 50 pA) and (d) EG (V$_{sample}$ = -0.1 V, I$_{tunneling}$ = 0.4 nA) demonstrating aligned lattice orientations.

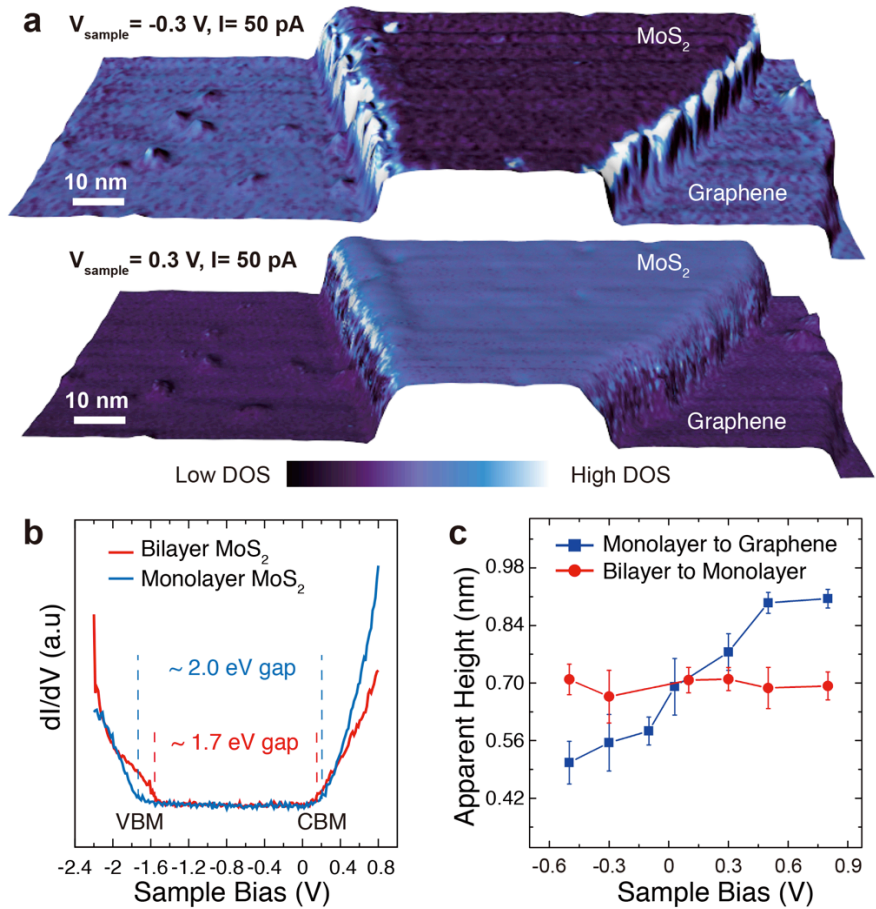

**Figure 4.** Electronic structure of MoS$_2$/EG. (a) STS mappings overlaid on three-dimensionally rendered topography images of MoS$_2$/EG at sample biases of -0.3 V and 0.3 V showing different relative DOS between MoS$_2$ and EG. (b) STS spectra taken on monolayer and bilayer MoS$_2$ showing a smaller bandgap of ~1.7 eV for bilayer MoS$_2$. (c) Bias-dependent apparent thickness of monolayer MoS$_2$ on EG, and apparent height of a monolayer to bilayer MoS$_2$ step edge.

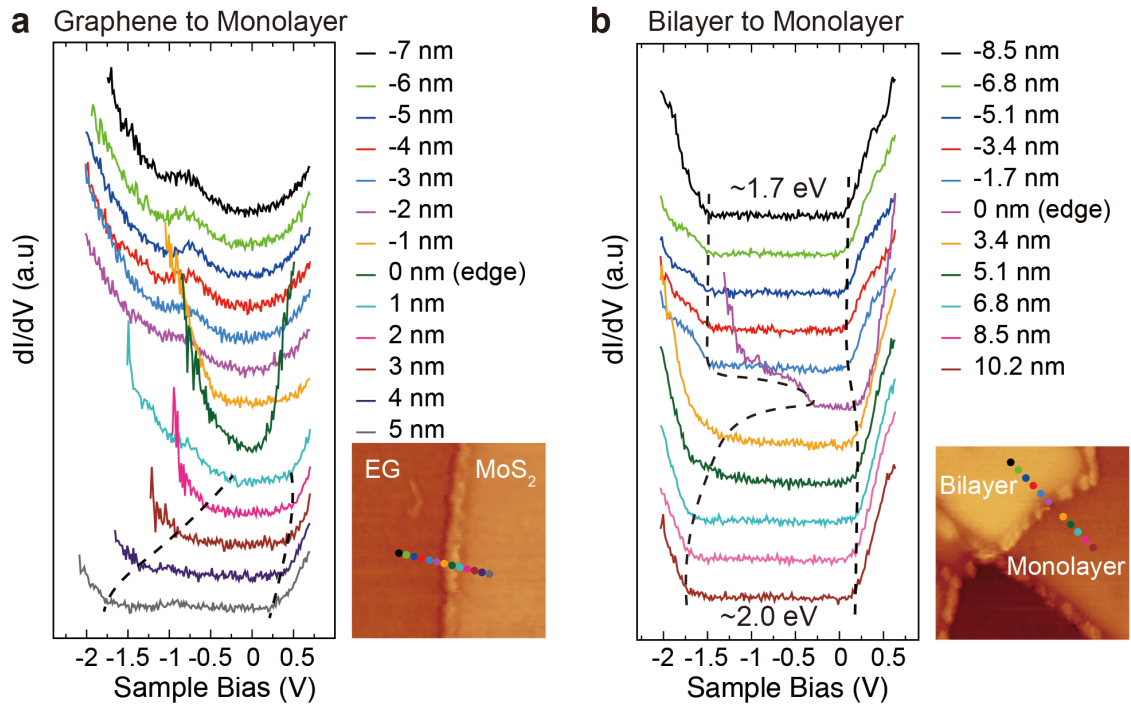

**Figure 5.** STS line scans across MoS$_2$ edges. (a) STS spectra taken across a monolayer MoS$_2$ edge from EG. (b) STS spectra taken across a step edge from bilayer to monolayer MoS$_2$ showing band bending and decreased bandgap near the edge to ~0.4 eV.

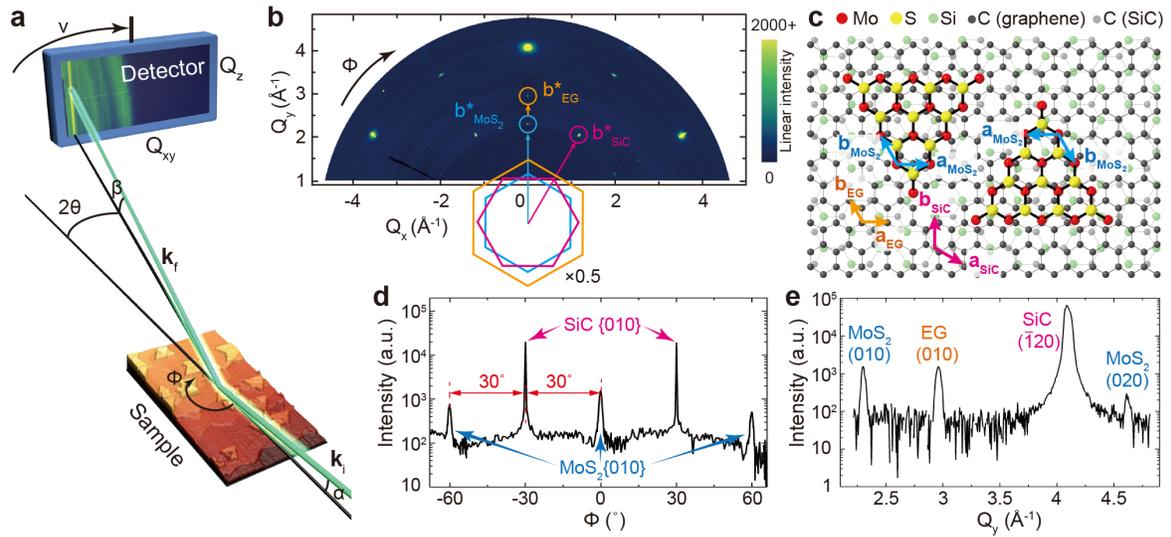

**Figure 6.** Rotationally commensurate van der Waals epitaxy of $MoS_2$ on EG. (a) Schematic of GIWAXS measurement. $\mathbf{k}_i$: incident wave vector, $\mathbf{k}_f$: scattering wave vector, α: incident angle, β: out-of-plane angle, 2θ: in-plane angle, ϕ: sample rotation angle, ν: detector rotation angle. (b) $Q_{xy}$ 2D reciprocal space map of $MoS_2$/EG projected from $Q_z = 0.08$ Å$^{-1}$ to $0.12$ Å$^{-1}$ by synchrotron GIWAXS. The **b\*** reciprocal space vectors are indicated by arrows. (c) Real-space model of the $MoS_2$/EG heterostructure with $MoS_2$ lattice aligned with that of EG. (d) Projected first order peaks of $MoS_2$ and SiC onto ϕ showing sharp distributions. (e) Referring to (b), in-plane scattered intensity along $Q_y$ direction at $Q_x = 0$. The determined real-space lattice constants of $MoS_2$, EG, and SiC are $3.16 \pm 0.01$ Å, $2.46 \pm 0.01$ Å, and $3.07 \pm 0.01$ Å, respectively.

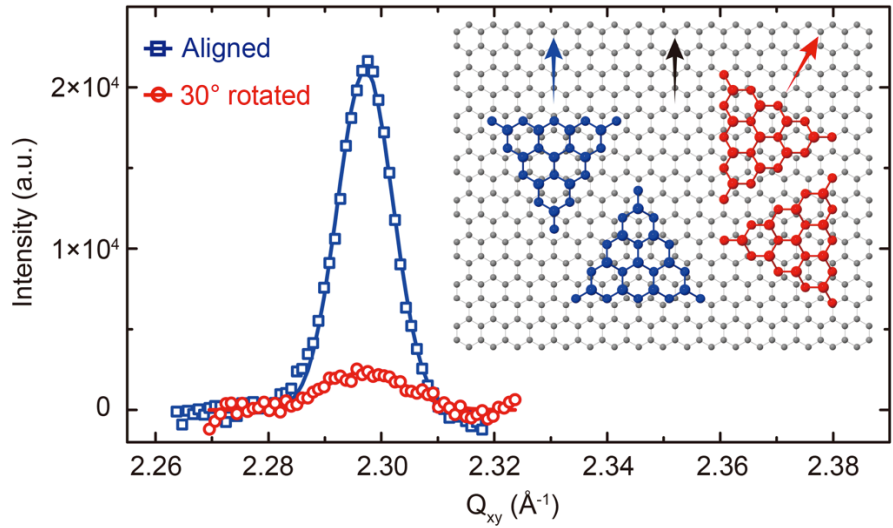

**Figure 7.** Examination of the growth orientation of MoS$_2$. GIWAXS data of the MoS$_2$(010) peak taken along the EG[010] direction (aligned growth, blue) and the SiC[010] direction (30° rotated growth, red). The integrated intensities of these two peaks are 2.6 × 10$^5$ and 4.2 × 10$^4$, respectively. Inset: Schematic of the two growth orientations of MoS$_2$ on EG. The armchair directions of MoS$_2$ from aligned (blue) and 30° rotated (red) growth are indicated by the blue and red arrows, respectively. The black arrow indicates the armchair direction of EG.